\documentclass[iop]{emulateapj}
\usepackage{graphicx}
\usepackage{amsmath}
\usepackage{natbib}
\usepackage{color}

\newcommand{\pgj}[1]{ #1}

\newcommand\new[1]{#1}

\newcommand{\lta}{\lesssim}

 \newcommand{\result} {\ifmmode{\pm4.5} \else {$\pm4.5$} \fi }
\newcommand{\mresult} {\ifmmode{\pm 19}  \else {$\pm19$} \fi }

\newcommand\figsco{
\begin{figure}[ht]
\begin{center}
{\includegraphics[width=1.0\linewidth]{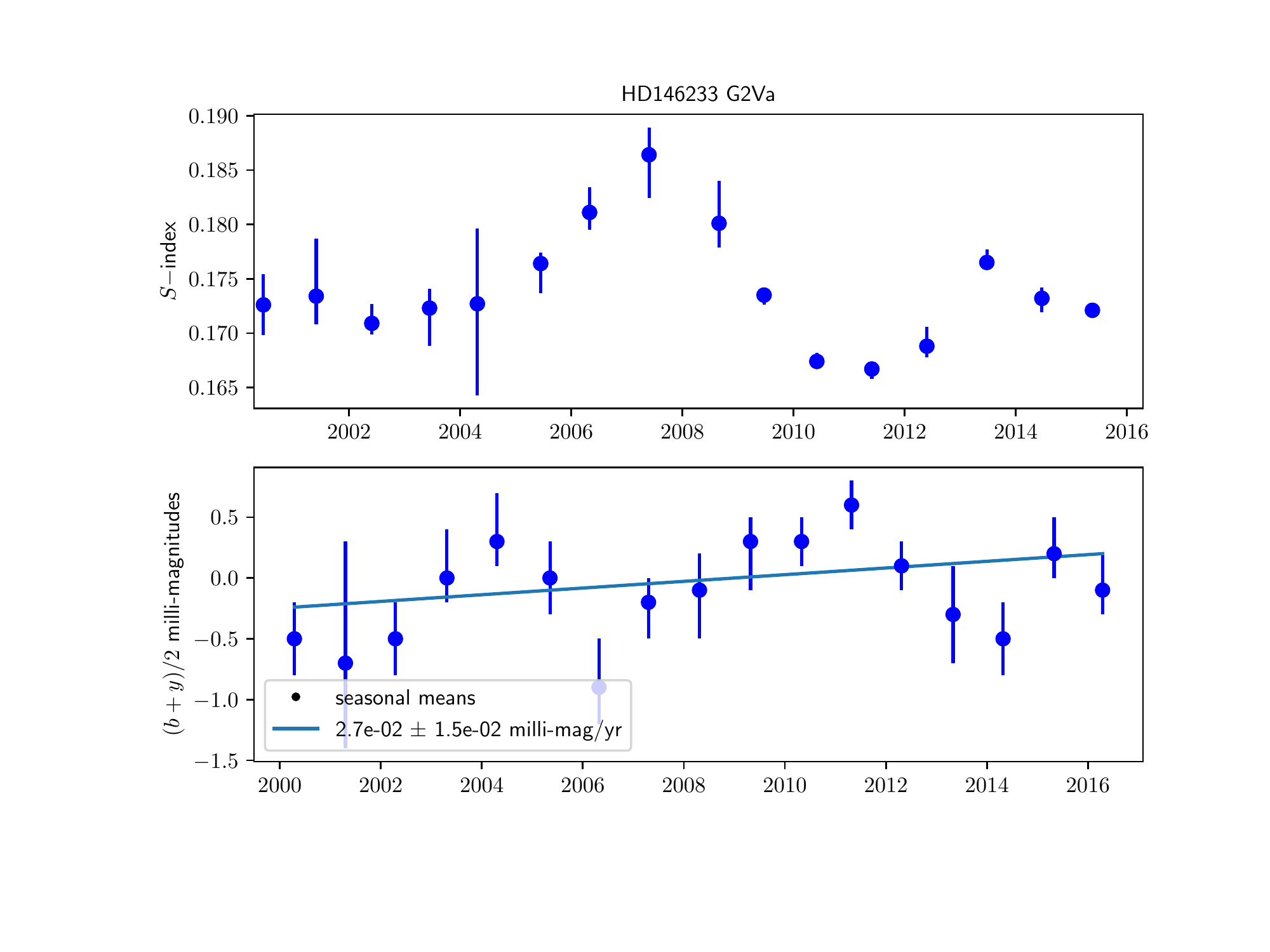}}
\end{center}
\caption{
Seasonally averaged data for the 
\ion{Ca}{2} ``S-index" and for 
the average magnitudes of the 
Str\"omgren $b$ plus $y$ filters 
are shown for the star 18 Sco (HD146233) whose properties are closest to those of the Sun.  
The straight line in the lower 
panel shows a least-squares fit
of a linear function to the photometric data, with the uncertainties shown
as listed in \cite{2018ApJ...855...75R}.  The
gradient is $28\pm 15$ micro-magnitudes per year.
}
\label{fig:18sco}  
\end{figure}
}

\newcommand\figslope{
\begin{figure}
\begin{center}
\includegraphics[width=\linewidth]{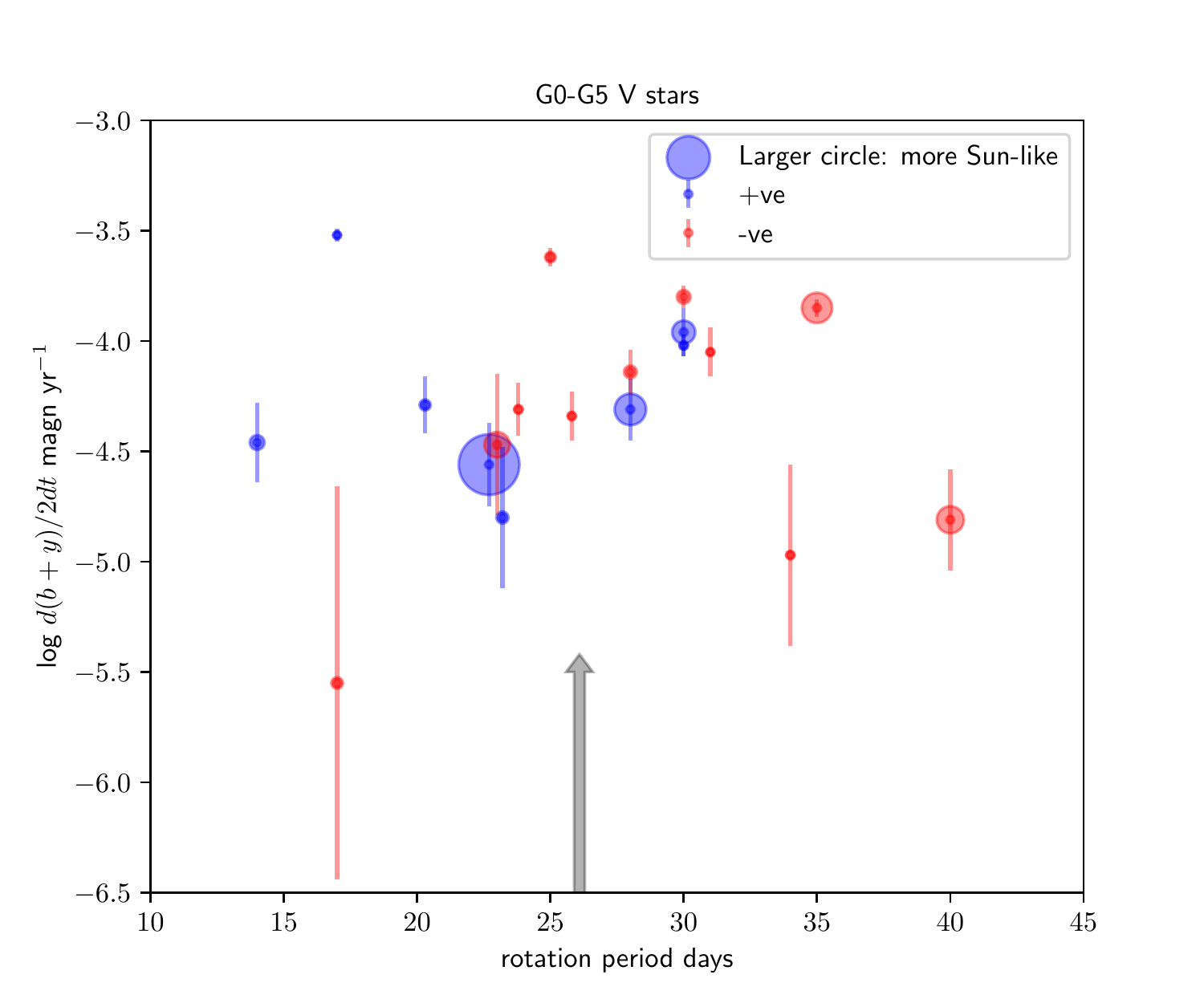}
\end{center}
\caption{
Computed gradients of the 
seasonally-averaged 
$(b+y)/2$ photometric measurements of \cite{2018ApJ...855...75R} 
are plotted as a function of 
stellar rotation period. 
\new{The ordinate is the logarithm
of the magnitude \textit{change}, i.e.
$\log_{10} \left(1.086 \log_{10} \Delta F/F
\right)$ where $F$ is the flux ($2.5\log_{10}e = 1.086$).   Thus, for  $\Delta F / F \ll 1$, the ordinate is proportional to 
$\log_{10} \Delta F$, i.e. the logarithm of the flux changes. }
The sizes
of the symbols are \new{inversely} proportional 
to the distance metric, the larger
the symbol, the \new{more} Sun-like is
the  star \citep{2018ApJ...855...75R}. 
The solar rotation period is marked with an arrow.}
\label{fig:slope}  
\end{figure}
}

\newcommand\figbpy{
\begin{figure}
{\includegraphics[width=1.0\linewidth]{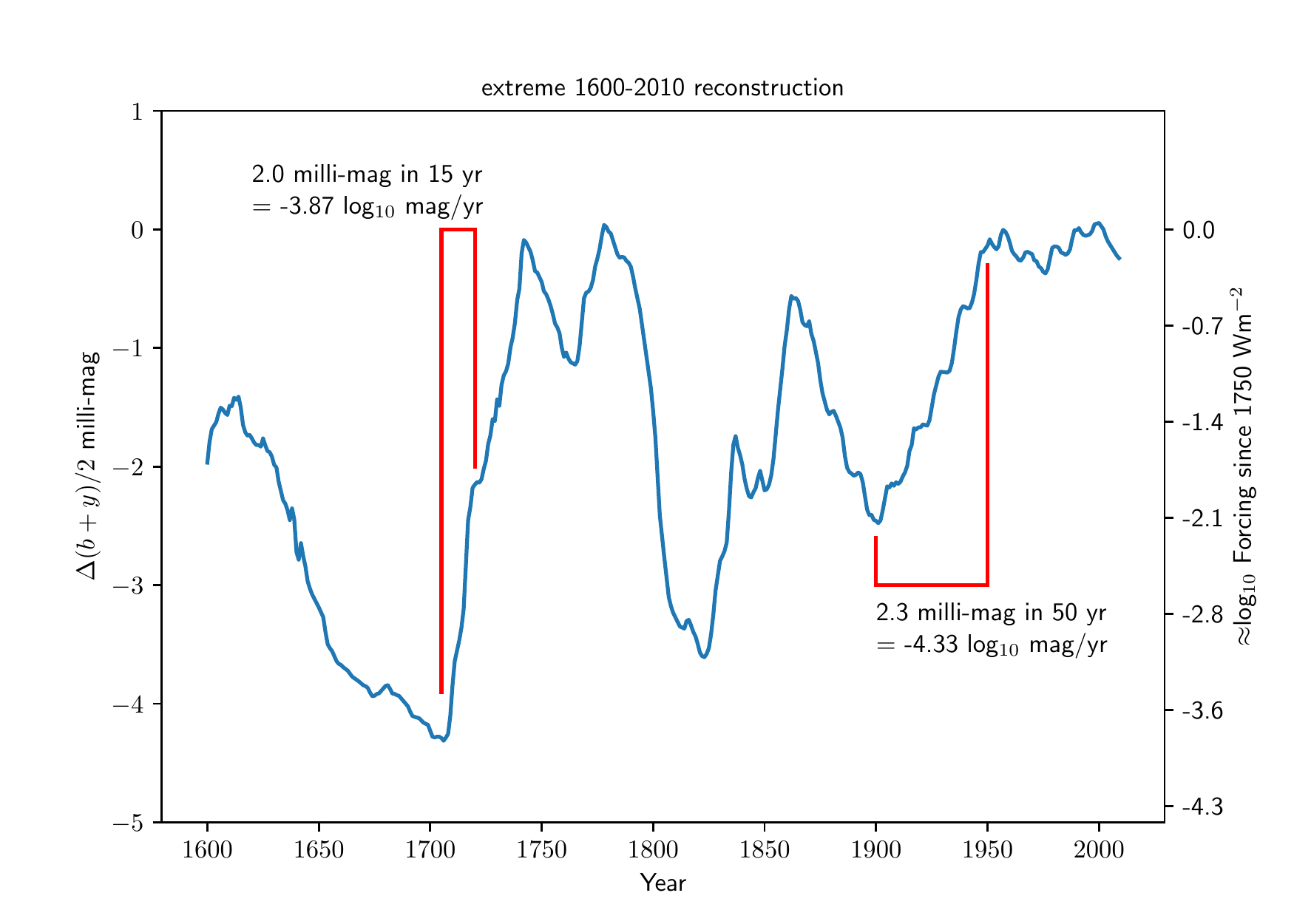}}
\caption{
Variations from a model by 
\cite{Shapiro+others2011} are shown, in which variations in radiative forcing since
1750 are of order 3 W~m$^{-2}$,
far larger than the IPCC estimate
of –0.30 to +0.10  W~m$^{-2}$
\cite{ar5}. 
(The relationship between Str\"omgren  magnitudes and 
irradiance
is linear, taken from
the model computations. The factor differs
from the values adopted in the text, but it is of no consequence for
the arguments in the text).
}
\label{bpy}  
\end{figure}
}

\newcommand\fighist{
\begin{figure}
{\includegraphics[width=1.0\linewidth]{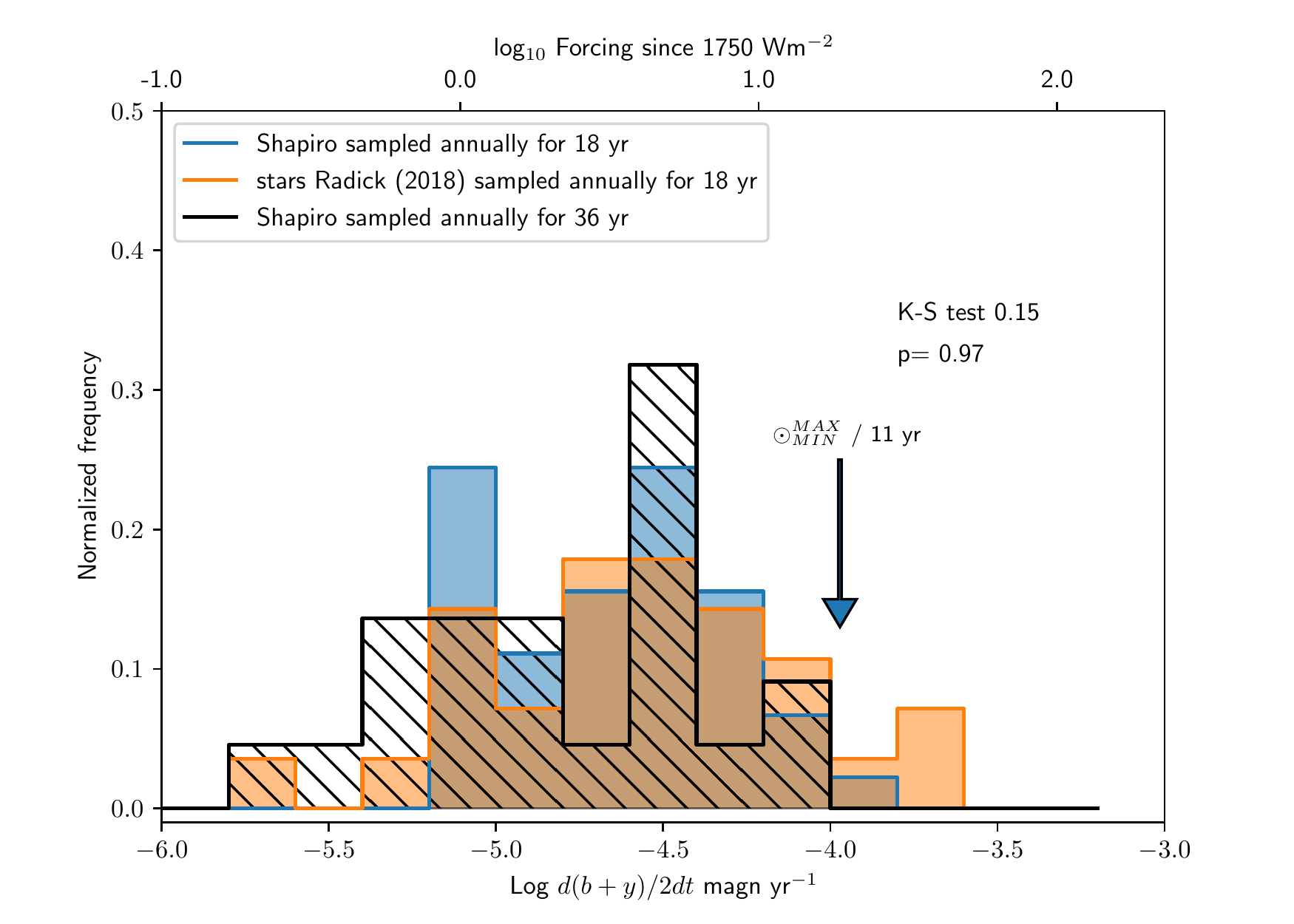}}
\caption{
The distributions of stars 
according to their derived 
secular slope ($d(b+y)/2dt$) 
is shown along with distributions
drawn from time-series extracted 
from the computations.   All three distributions are compatible with being drawn from 
the same underlying distribution
according to the Kolmogorov-Smirnoff test. Statistics are shown for the test applied to the two 18-year distributions. 
(The first parameter is a measure of non-parametric ``distance'' between two distributions, the second is the probability that they are drawn from the same distribution).   Notice that the 36-year calculated distribution 
moves to the left by about a factor of two.
}
\label{fig:hist}  
\end{figure}
}

\newcommand\figdur{
\begin{figure}
\begin{center}
\includegraphics[width=1.0\linewidth]{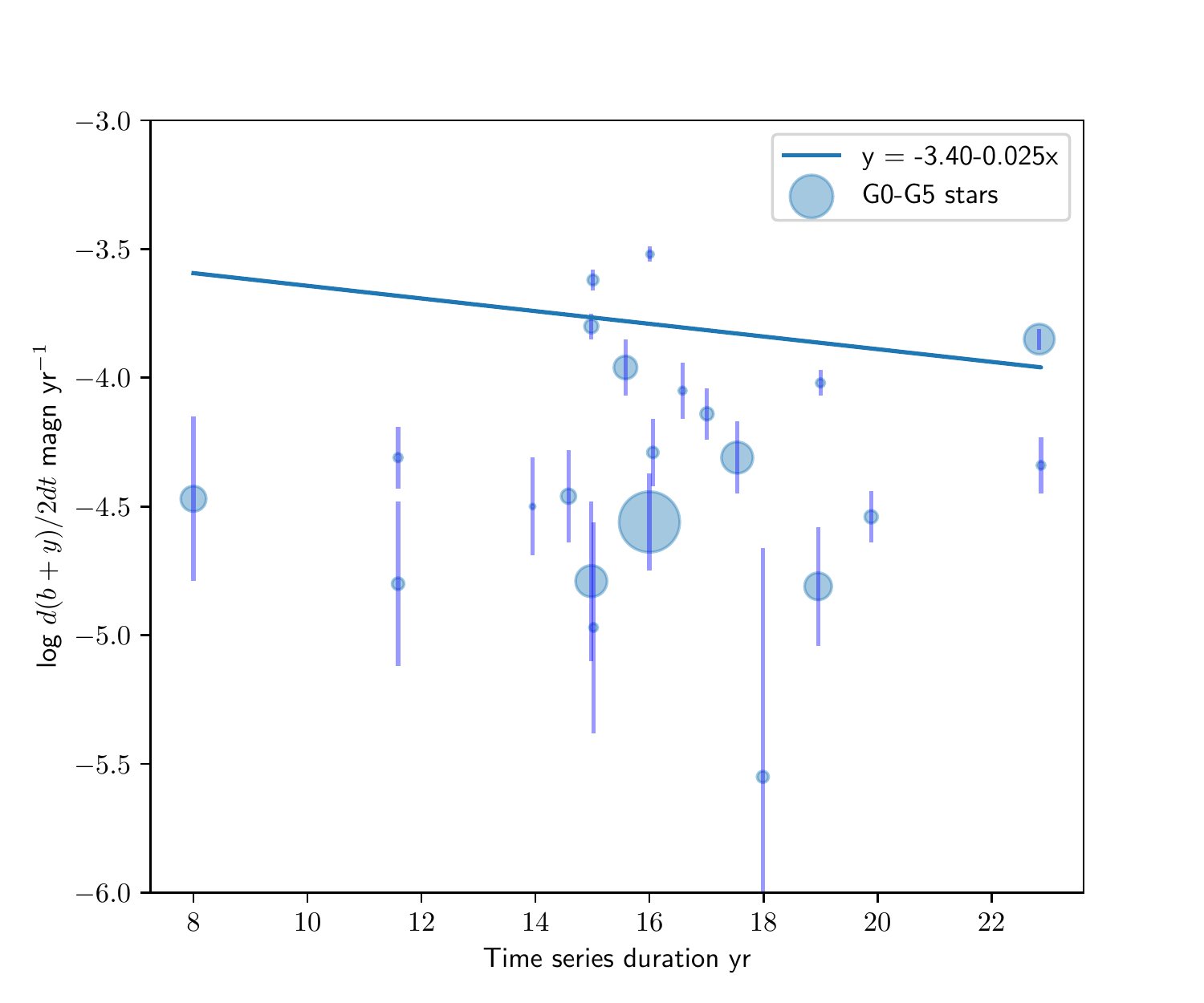}
\end{center}
\caption{
Secular gradients are 
shown as a function of time series duration.  \new{The solid line shows a 
linear least-squares fit to the data taking into account the 
formal error estimates to the fits for each star}. 
}
\label{fig:durv}  
\end{figure}
}

\newcommand\figfast{
\begin{figure}[ht]
\begin{center}
{\includegraphics[width=1.0\linewidth]
{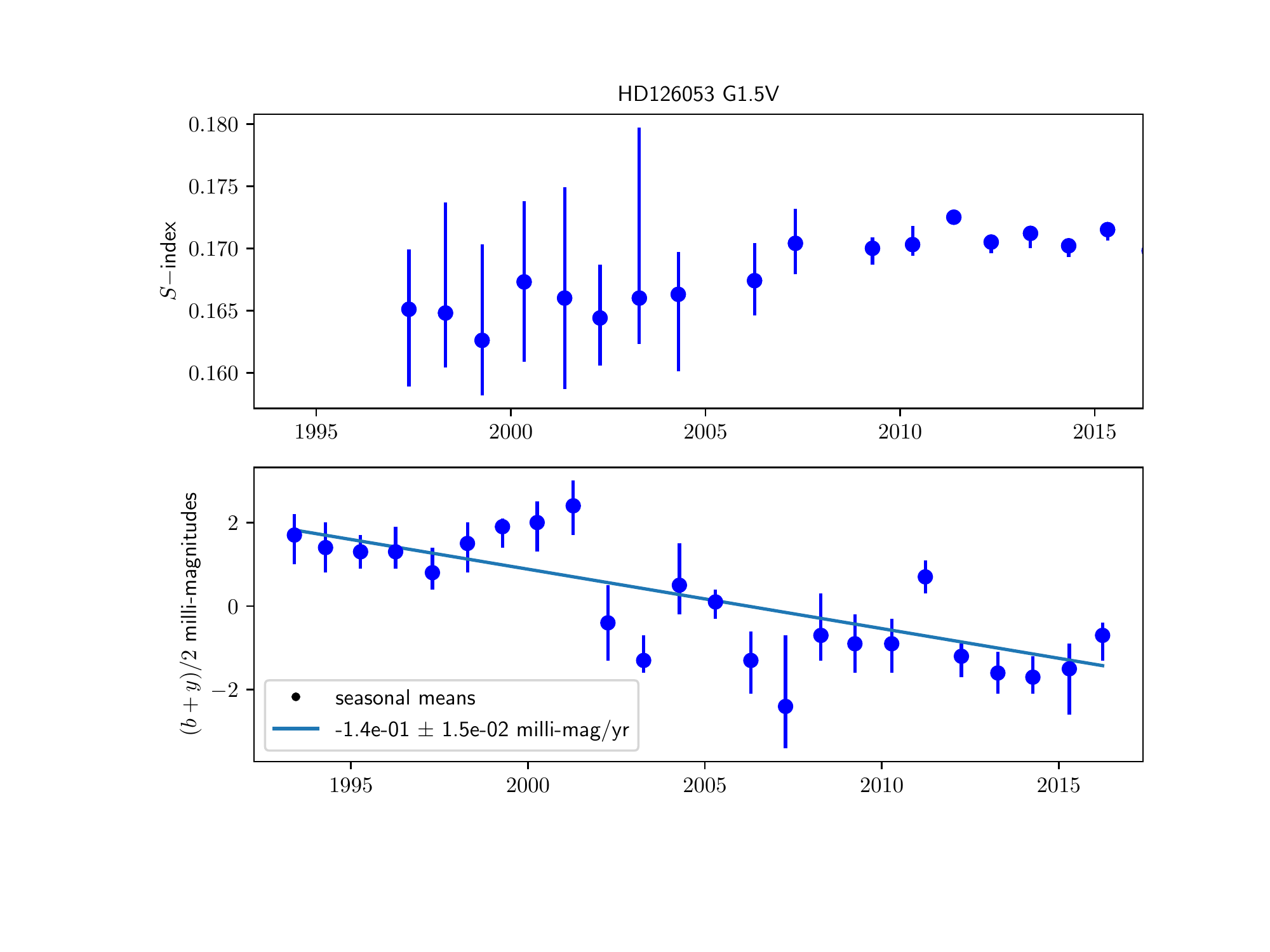}}
\end{center}
\caption{Data are shown 
as in Figure~\ref{fig:18sco}
for the star HD126053.  This star 
occupies the point (23,-3.8) in Figure~\ref{fig:durv}.
It has a well-defined gradient and is very similar
to the Sun with a value of $dis=0.26$. 
The
gradient is $140\pm 15$ micro-magnitudes per year, five times that of 18 Sco (Figure~\ref{fig:18sco}).
}
\label{fig:fast}  
\end{figure}
\vskip 15pt
}

\begin{document}


\title{Sun-like stars shed light on solar climate forcing}

\author{P. G. Judge, R. Egeland}
\affil{High Altitude Observatory, National
Center for Atmospheric Research,\\
P.O. Box 3000,
Boulder CO~80307-3000, USA. }
\and
\author{G. W.  Henry}
\affil{Center of Excellence in Information Systems,\\ Tennessee State University, Nashville, TN 37209, USA}

\begin{abstract}
Recently published, precise stellar photometry of 72 Sun-like stars obtained at the Fairborn Observatory 
between 1993 and 2017 is used to set limits on the solar forcing of Earth's atmosphere of
\result{}  W~m$^{-2}$ since 1750. 
This compares with
the $+2.2\pm1.1$ W~m$^{-2}$ IPCC estimate for anthropogenic forcing.  Three critical assumptions are made.
In decreasing
order of importance they are:  (a) most of the brightness variations occur within the average time-series length of
$\approx17$ years; (b) the Sun seen from the ecliptic behaves as an 
ensemble of middle-aged solar-like stars; and 
(c) narrow-band photometry in the Str\"omgren $b$ and $y$ bands are linearly proportional to the total
solar irradiance.   Assumption (a) can best be relaxed and tested by obtaining more
photometric data of Sun-like stars, especially those already observed.  Eight stars with near-solar
parameters have been observed from 1999, and two since 1993.  Our work reveals the importance of continuing
and expanding ground-based photometry, to complement expensive solar irradiance measurements from space. 
\end{abstract}


\maketitle


\section{INTRODUCTION}

The influence of the variable total solar irradiance of Earth (TSI) has remained a major uncertainty in our
ability to predict quantitatively how the Sun might contribute to climate change
\citep[e.g.,][]{Lean2018,Dudok2018}.  Research in this area is active, but is notoriously plagued by
difficulties including historically inaccurate (but precise) irradiance measurements from space, the use of
extrapolations based upon linear ``proxies", problems of interpreting incomplete stellar \pgj{datasets, and the
general lack of accurate long-term ($>$ decade-long) variability data of the Sun and stars, {problems
eloquently summarized by \citet{2011GeoRL..38.6701S}.  For example, the recently measured differences
between the last sunspot minimum of 2008 and earlier minima have sparked much debate, new propositions, 
and further speculation about future and past solar behavior 
\citep[e.g.][]{2011GeoRL..38.6701S, 2013JAdR....4..209H}}}.

In this article we use precise stellar photometry over the past quarter century \citep{2018ApJ...855...75R} to
set limits on the rate at which the Sun might vary over the next few decades.  This new approach takes
advantage of these multi-decade data to set statistical limits on the variability of Sun-like stars.  To
proceed, we must make several assumptions.  The important assumptions are:  (a) variances
(integral of power spectra over all frequencies) are not much larger than those sampled over the average
time series durations of 17 years; (b)  the Sun behaves in a fashion represented by a carefully
selected stellar ensemble as observed from Earth, noting that solar radiation is received at Earth in the ecliptic plane, which is tilted just 7$^o$ from  the solar equatorial plane; and (c) that the average of the Str\"omgren $b$ and $y$ filter differential
magnitudes is linearly proportional to TSI.   

Assumption (b) has
been studied empirically by 
\citet{1993JGR....9818907S,2001A&A...376.1080K}, the more recent work suggesting that (b) is justified to
about 6\%{} levels. 
Assumption (c) is discussed in depth by \citet{2018ApJ...855...75R}.  All of these assumptions are testable with further
measurements and (perhaps) physical models.
\figsco
With these assumptions, our careful assessment of uncertainties, along with consistency checks of the stellar time series, we estimate a limit to the secular change of \mresult{} milli-magnitudes (0.019 mag) of change in brightness over the standard period
of 250 years \citep{ar5}.  This amounts to a forcing of \result{} W~m$^{-2}$ since 1750, and some five times
smaller over the next 5 decades.

\section{DATA SELECTION AND ANALYSIS}

The data analyzed here come primarily from two sources.  First, the most precise and stable set of photometric
measurements of Sun-like stars (SLS) has been painstakingly acquired by one of us (G.W.H.) using robotic
telescopes designed, constructed, and maintained by Louis Boyd at Fairborn Observatory.  The photometric data,
covering up to 24 years between 1993 and 2017, have been processed and vetted mainly by G.W.H. 
\citep[described in][]{1999PASP..111..845H}.  The Fairborn data were recently published by
\cite{2018ApJ...855...75R} and made freely available.  The second source of data is the Lowell Observatory
program on solar and stellar chromospheric activity that produced time series of the magnetically sensitive
\ion{Ca}{2} line strengths between 1992 and 2016.  These measurements were converted to the physical
parameter $R'_{\mathrm HK}$, the ratio of flux in the \ion{Ca}{2} lines relative to the stellar luminosity.  
As a cooperative program with Fairborn, the Lowell observations of the same stars were published together with
the photometric results in \cite{2018ApJ...855...75R}.  We augmented these data with rotation periods and
Rossby numbers of these same stars from Table 5.5 of Egeland's PhD thesis \citep{2017PhDT.........3E}.
\pgj{Further refinement of Egeland's carefully vetted data were needed to identify true SLS by ensuring
consistency with a robust rotation/activity/age relation \citep{2008ApJ...687.1264M}.  Only two stars 
(HD 86728 and HD 168009) were thus rejected from further analysis.  Their periods are from sparse time series 
of \ion{Ca}{2} data over one season from \citealp{2016A&A...586A..14H} that seem to us to be overtones of the
rotation period.  For other stars of particular interest, owing to their similarity to the Sun, we used the
rotation/age relationship to estimate ages/rotation rates, and hence the Rossby number $Ro$ (see below and 
Table 1).  These data were used to reject stars on the basis of their different ages, activity levels, and
variability.} 

We initially examined the time series of all 72 stars of \cite{2018ApJ...855...75R} without reference to
stellar age, elemental abundances, gravity, effective temperature and other parameters.  These data 
are ideally suited to time-series analysis.  All the necessary processing, vetting and calibrations have
been done.  Unbiased (seasonal) averages of photometric brightness in the standard Str\"omgren $b$ and $y$
filters were derived, uncertainties quantified, and consistency checks carefully made.  Data for our star
most similar to the Sun (18 Sco) are shown in Figure~\ref{fig:18sco}.  It should be noted that each data 
point for each year consists of many individual measurements, with attention given to a proper quantification 
of all uncertainties, including those from variations in comparison stars \citep{2018ApJ...855...75R}. 

We seek limits on secular (not cyclical) changes in stellar brightness.  Therefore, for each star $i$, 
the gradient of the time series and its formal uncertainty were obtained as shown in
Figure~\ref{fig:18sco}, and saved as $g_i\pm \sigma_i$, $i=1\ldots 72$.  We then derived the ensemble mean
gradient $\langle G(\tau) \rangle$ and its uncertainty $\sigma(\tau)$.  Each $g_i$ has associated with it
the time series duration $\tau_i$ from which it was so-derived; $\tau=17$ years is the mean duration of
the stellar time series.  Cyclical variations that occur in roughly 30\%  of SLS
\citep{2017PhDT.........3E} will naturally contribute to the gradients derived, depending on the amplitude,
phase and duration of the cycles. Longer time series will of course reduce the derived gradients of such
stars.
\figslope

Now we invoke the ergodic hypothesis, i.e. that the Sun's brightness variations in time are
\textit{statistically} identical to a random sample of SLS (defined below) over a time
scale $\tau$ of $17$ years.  We can then interpret $\langle G(\tau) \rangle \pm \sigma(\tau)$ as the
magnitude of changes in solar brightness averaged over any given epoch covering any contiguous $\tau$
years.  The figure of interest here is \textit{not} $\langle G(\tau) \rangle$ itself of course, but
$\sigma(\tau)$.

By invoking this hypothesis, we assume that essentially all of the variance in brightness of SLS occurs
within the $\lta 17$-year span $\tau_i$ of the stellar observations.  The value of 
$\langle G(\tau) \rangle \pm \sigma(\tau)$ so-derived can be strictly applied only to solar data for time
spans $\lta \tau$.  If our strong assumption (a) later turns out to be true, then we can extend this
strict limitation to longer periods, for example enabling us to estimate variations in solar TSI since
1750. 

The sample of 72 stars was winnowed down on the basis of the ``metric'' measuring the distance of a 
given star from the Sun defined by \cite{2018ApJ...855...75R}, listed and 
described in our Table 2.  In addition, we required that each
star be of luminosity class V and have a well-determined measure of activity (we examined rotation period,
Rossby number Ro, age, and $R^\prime_{HK}$) from which a more ``Sun-like" set of stars was found.  The
stars that survived all criteria for selection are listed in Table 1. 
\begin{deluxetable}{rlllccl}
\tablecaption{ \label{tab:stars} The subset of 
stars analyzed}
\tablehead{
HD & Sp. & B-V & \multispan{2} age~(Gyr) & $p_{rot}$ & var.\\ 
 &  &  & low & up & days & type}
\startdata 
$^\ast$1461 & G3VFe0.5 & 0.68 & 0.9 & 3.1 & 17.0 & poor \\
10307 & G1V & 0.62 & 3.5 & 8.2 &  &  \\
13043 & G2V & 0.62 & 4.3 & 7.6 & 34.0 &  \\
$^\ast$38858 & G2V & 0.64 & 3.2 & 7.5 & 40.0 &  \\
42618 & G4V & 0.64 &  &  &  &  \\
43587 & G0V & 0.61 & 4.45 & 5.49 & 20.3 & flat \\
$^\ast$50692 & G0V & 0.6 & 4.0 & 6.0 & 25.0 &  \\
$^\ast$52711 & G0V & 0.59 & 4.9 & 9.7 & 30.0 &  \\
$^\ast$95128 & G1-VFe-0.5 & 0.61 & 6.03 & 6.03 & 30.0 &  \\
$^\ast$101364 & G5 & 0.65 & 3.5 & 3.5 & 23.0 &  \\
109358 & G0V & 0.58 & 5.3 & 7.1 & 28.0 &  \\
120066 & G0V & 0.59 &  &  &  &  \\
126053 & G1.5V & 0.63 & 5.49 & 5.49 & 35.0 & poor? \\
141004 & G0-V & 0.6 & 5.8 & 6.7 & 25.8 & long \\
143761 & G0+VaFe & 0.6 & 8.5 & 11.9 & 17.0 & long \\
146233 & G2Va & 0.65 & 3.65 & 3.75 & 22.7 & good \\
$^\ast$157214 & G0V & 0.62 & 4.1 & 6.6 & 14.0 & irr. \\
$^\ast$159222 & G1V & 0.62 & 3.5 & 6.0 & 28.0 &  \\
$^\ast$186408 & G1.5Vb & 0.62 & 6.7 & 7.3 & 23.8 & flat \\
$^\ast$186427 & G3V & 0.66 & 6.7 & 7.3 & 23.2 & flat \\
$^\ast$187923 & G0V & 0.65 & 8.1 & 9.5 & 31.0 &  \\
$^\ast$197076 & G5V & 0.61 & 0.2 & 9.3 & 30.0 & 
\enddata
\tablecomments{Upper and lower limit estimates of
stellar ages are listed under ``low'' and ``up'' in Gyr.
The ages are from \citet{2017PhDT.........3E}, except where marked
with an asterisk, where ages are cruder estimates  from 
isochrones in the literature, using \textit{HIPPARCOS} distances and visible magnitudes.  For these stars the rotation-age
relations were  used to estimate 
$p_{rot}$ and Ro (Table 2) except when 
rotation periods were known.  }
\end{deluxetable}

Fortunately, the results depend little on the precise choice of selection parameters.  The best result with
the smallest dispersion of gradients was found by restricting the sample to stars with the ``activity
parameter" $\log R^\prime_{HK} \le -4.8$, which is close to the solar value of ${-4.94}$.  This final
restriction yielded a sample of 22 stars with an ensemble mean gradient  
\begin{equation}
    \langle G(\tau) \rangle \approx
  -6\mresult
   {\rm\  micro-magnitudes\ per\ year.}
\end{equation}
The estimate is consistent with a value of zero, as it must be if a large enough number of stars behave 
independently.  The gradients derived from this set of stellar time series are shown in
Figure~\ref{fig:slope}, plotted as a function of stellar rotation period.  The linear trends extracted are
given in Table 2.
\figdur

The ensemble mean gradient corresponds to a forcing of the climate by solar irradiation alone of 
\begin{equation}
    \Delta F(\tau)_\odot \approx 
   -1.5\result  {\rm\  W~m^{-2}~since~1750},
\end{equation}
where we have used $G = (1.55 \pm 0.37) \Delta F/F$ to convert from milli-magnitudes to irradiance changes
$\Delta F$ in W~m$^{-2}$ \citep{2018ApJ...855...75R} for an average irradiance of $F=1361$ W~m$^{-2}$. 
\pgj{The important figure here is the range of the slope from the uncertainties of \result{} W~m$^{-2}$.}
The significance of this estimate is seen when compared with the climate forcing since 1750 due to
anthropogenic effects, which is estimated by the IPCC \citep{ar5} to be 
\begin{equation}
    \Delta F_{AG} \approx 1.1\  \mathrm{to}\  3.3
     {\rm\  W~m^{-2}~since~1750}.
\end{equation}
\begin{deluxetable}{rlllcccc}
\tablecaption{ \label{tab:starsd} Derived stellar  properties}
\tablehead{
HD & dis. & Ro & $\log R^\prime_{HK}$ & $\tau$ &
\multispan{3} ------------Gradient------------ \\
&&&&& $\log g_i(\tau)$  & sign & $\log \sigma_i$ \\
 &  &  & & yr & mag/yr & & mag/yr
 }
\startdata 
1461 & 0.68 & 1.8 & -5.04 & 17.99 & -5.55 & $-$ &  0.89 \\
10307 & 0.61 &  & -5.01 & 19.89 & -4.54 & $+$ &  0.10 \\
13043 & 0.92 & 3.3 & -5.01 & 15.02 & -4.97 & $-$ &  0.41 \\
38858 & 0.29 & 3.5 & -4.89 & 18.96 & -4.81 & $-$ &  0.23 \\
42618 & 0.25 &  & -4.96 & 14.98 & -4.79 & $-$ &  0.31 \\
43587 & 0.71 & 2.6 & -4.99 & 16.06 & -4.29 & $+$ &  0.13 \\
50692 & 0.74 & 2.84 & -4.96 & 15.01 & -3.62 & $-$ &  0.04 \\
52711 & 0.57 & 3.6 & -4.96 & 14.98 & -3.80 & $-$ &  0.05 \\
95128 & 0.89 & 3.0 & -5.06 & 19.0 & -4.02 & $+$ &  0.05 \\
101364 & 0.31 & 1.9 & -4.97 & 8.0 & -4.47 & $-$ &  0.32 \\
109358 & 0.61 & 4.9 & -4.97 & 17.01 & -4.14 & $-$ &  0.10 \\
120066 & 1.47 &  & -5.14 & 13.95 & -4.50 & $+$ &  0.19 \\
126053 & 0.26 & 3.04 & -4.94 & 22.84 & -3.85 & $-$ &  0.04 \\
141004 & 0.93 & 2.84 & -4.97 & 22.87 & -4.34 & $-$ &  0.11 \\
143761 & 1.08 & 1.87 & -5.09 & 16.01 & -3.52 & $+$ &  0.03 \\
146233 & 0.13 & 1.9 & -4.93 & 16.0 & -4.56 & $+$ &  0.19 \\
157214 & 0.53 & 1.37 & -5.01 & 14.58 & -4.46 & $+$ &  0.18 \\
159222 & 0.25 & 2.7 & -4.89 & 17.54 & -4.31 & $+$ &  0.14 \\
186408 & 0.88 & 1.89 & -5.07 & 11.59 & -4.31 & $-$ &  0.12 \\
186427 & 0.65 & 1.9 & -5.04 & 11.59 & -4.80 & $+$ &  0.32 \\
187923 & 1.01 & 2.6 & -5.05 & 16.58 & -4.05 & $-$ &  0.11 \\
197076 & 0.34 & 3.0 & -4.89 & 15.58 & -3.96 & $+$ &  0.11
\enddata
\tablecomments{``dis." is the measure of 
dissimilarity of the star from the Sun 
\citep{2018ApJ...855...75R}. It is defined  
 by measuring its distance from the Sun in a three-dimensional $M_V$, $B-V$ and , $\log R^\prime_{HK}$ manifold.
}
\end{deluxetable}

The stars are therefore tantalizingly close to providing useful constraints on magnetically-induced solar
irradiance variations, \textit{independent of any other measurements or assumptions.} 

On face value, the uncertainties and shortness of the time series of SLS limit the apparent usefulness of 
stellar photometry in addressing pressing climate change problems facing humanity \citep{ar5}.  However,
the present work represents only the first measurements to limit the irradiances of SLS on
periods that otherwise require \textit{untestable extrapolations (``reconstructions'') or the patching
together of different satellite measurements} of total solar irradiance by ad-hoc offsets in radiometric
calibrations.   

The current IPCC estimates of solar forcing (-0.3 to +0.1 W~m$^{-2}$) \citep{ar5} are an order of magnitude
smaller.  However, these numbers have been derived using precisely those extrapolations based upon
``proxies'' that we are specifically trying to avoid.  They are more educated guesses than hard data.  

It therefore is important to see how stellar photometry might yield improved results through longer data
sets.   
\begin{itemize}
    \item Observing stars over a longer time span will measure more of the low-frequency $(1/\tau_i)$ 
    components of the power spectrum that contribute to the variances in brightness.  
    \item Depending on the (unknown) amount of power at low frequencies, the increase in lengths of time
    series may or may not decrease the variances of the measured gradients.  In the limit where all the
    power has been captured in $\tau=17$ years, the slopes and their standard deviations will vary roughly as 
    $1/\tau_i$.  
    \item  Observing a larger number $N$ of stars will reduce the uncertainties by a factor $\sqrt{N}$. 
\end{itemize}

The second point is illustrated by comparing the statistical stellar behavior against 
``a reconstruction'' with a very large irradiance variation (several W~m$^{-2}$) since 1750.  We examine
this below (Section \ref{sec:rec}).  An example of a large linear trend is shown in Figure~\ref{fig:fast},
showing data for HD 126053, which occupies the point near (23,-3.8) in Figure~\ref{fig:durv}.  The star is
very similar to the Sun ($dis=0.26$); it has been observed for 22 years; yet it has a trend 5 times that of
18 Sco (Figure~\ref{fig:18sco}).
\figfast
But the essential implication of the length of time series  must be important, because a plot of linear trend against
length of time series $t_{dur}$ for the sample gives 
$$
\frac{1}{2}\frac{d}{dt} (b+y) = -3.40
-0.50 \frac{t_{dur}}{20}
$$
with $t_{dur}$ in years (Figure~\ref{fig:durv}).

Additionally, earlier 
photometric data of one target 
(HD143761) were published by \citet{1997ApJ...485..789L}. While these were obtained with 
a different system at Lowell Observatory having
larger uncertainties
than those of \citet{2018ApJ...855...75R}, they were  compared with the same 
standard star. By assuming that the 
average of each time series for HD143761 are identical (again, a strong assumption), we can effectively
extend the time series from 
16 to 32 years (1984-2016). Under the
strong assumption, the gradient 
is reduced from -3.52 (Table 2) to -3.92. 
Therefore, we can reasonably expect the gradients  to decrease
with increasing time series duration.  

The last bulleted point has a few practical problems, given that society would like information on the role
of solar variations as a source of global warming or cooling in the next few decades.  First, new time
series would build up from year zero, and at least a decade would pass before meaningful statistics could
be derived.  Second, the selection of good comparison stars is a tedious but important problem, requiring
human vetting to achieve reliable results \cite[e.g.][]{1999PASP..111..845H}.  Lastly, the number of
genuinely SLS bright enough to measure with modest (meter-class) telescopes is small. As
measured by the number of stars in a meaningful volume of hyper-space similar to the Sun
\citep{1998ApJS..118..239R}, considerable work would be needed to identify new, dimmer targets.  

There remains the nagging question of whether the Sun is different from other SLS
\citep{1998sce..conf..419G}.  \citet{2018ApJ...855...75R} conclude:
\begin{quote}
    ``it may be unusual in two respects:  (1) its comparatively smooth, regular activity cycle, and (2) its
    rather low photometric brightness variation relative to its chromospheric activity level and
    variation\ldots''
\end{quote}

These authors speculate that facular brightening may nearly balance sunspot darkening, explaining the
second point.  \citet{2017PhDT.........3E} pointed out that the Sun has the most regular cycle of all
Sun-like stars measured so far.

The question of whether the Sun acts (magnetically) as other SLS is difficult to answer.  If all
such stars are indeed magnetically similar, it implies that stars have a consistent magnetic variability over time scales of several Gyr (the age range of our sample) to $\sim$100 million years. The latter is close to 
the uncertainty in ages of older main sequence stars obtained using the best available methods.  It is impossible to verify or refute the question for the Sun, even using a cosmogenic proxy record, which presently stretches only 0.01 million years into the past \citep{Wu2018}. Certainly, the most Sun-like of the stars found so far, 18 Sco (HD 146233) has clear differences in metallicity and
starspot cycle length.  Nevertheless, there is hope that a carefully selected stellar ensemble can represent
the activity of the Sun in middle and old age.  \citet{vanSaders2016} demonstrated that rotation rates of
middle-aged and old GV stars converge as a result of weakened magnetic breaking.  Unlike younger stars,
there is perhaps a good \textit{physical} reason to believe that magnetic dynamos of older Suns, and their effects, should be similar. 

\section{TIME SERIES FROM A SOLAR ``RECONSTRUCTION''}
\label{sec:rec}

The limits of our analysis due to the lack of longer time series can be illustrated through a comparison of
our results with a ``reconstruction'' of solar variations with extraordinary and significant forcing of 
$\approx 6$ W~m$^{-2}$ from 1600 to 2010 \citet{Shapiro+others2011}.  Figure~\ref{bpy} highlights two
extended periods of near-monotonic large changes predicted over 15 and 50 years.  The first is compatible
with several stars (HD 126053, 52711, 50692, 143761; Table~2).  The second (50-year) period is compatible
with about half of the stars listed.
\figbpy

A distribution of the number of stars of a given slope is compared with equivalent distributions 
extracted from the time series from the reconstruction model in Figure~\ref{fig:hist}.  
\pgj{The two distributions (and the third corresponding to a 36-year span of solar observations) are
statistically compatible with the same underlying distribution, according to the standard
Kolmogorov-Smirnoff test.  However, a peak near $10^{-4.5}$ magnitudes per year persists in the reconstructed 
distribution. The peak arises 
mostly from the periods of 
\textit{long-term} variations, two of which are highlighted 
in Figure~\ref{bpy}.

\fighist

Our comparison of an (albeit) extreme solar reconstruction with stellar data is a reminder that
precise photometry requires patience. It would be unfortunate if the precise photometry
performed since 1993 were not followed up with similar data over the next few decades to constrain further
long-term solar variability.}

\section{CONCLUSIONS}

Already we have measurements of stellar behavior over periods longer than any direct and stable measure of
solar irradiance.  (Of all
experiments, VIRGO on the SoHO
spacecraft has operated almost
continuously for 24 years, but 
it suffers from difficult 
calibration issues over this
period, see \citealp{2015A&ARv..24....3P}).
Our limit of \result{} W~m$^{-2}$ of solar forcing since 1750 hinges on two critical
assumptions:  first, that the Sun behaves like a member of an ensemble of SLS; second, that the
stellar sample has measured essentially all of the variance in the seasonal stellar time series, from a
frequency of $1/17$ years$^{-1}$ to 2 years (Nyqvist limit).  According to current understanding these
changes occur because of magnetic activity.  Certainly we can expect more power to be present on longer
time scales owing to magnetic variations among the stars.  But the question is, how much?  In this regard
we note that the length of time series is only 0.8 of the solar magnetic activity cycle.  Thus we might
expect some of the larger gradients to begin dropping out with additional data for those stars that are
known to be cycling (or perhaps irregular, see \citealp{2017PhDT.........3E}), as the linear trends become
replaced by cycles that might return to the same brightness, given two or more complete cycles.  

Only by observing these stars for longer periods can we set tighter limits on the ensemble's typical
behavior (Figure~\ref{fig:hist}). It is therefore of great importance to find a way to continue the
observational program pioneered at Fairborn Observatory.  \new{With the advent of remotely controlled automated telescopes, a cost-effective way to continue these measurements 
is surely within reach. The challenges
to obtain funding for such work 
remain to be addressed, as the Fairborn 
observatory work cannot continue for long
without investment in people as well as
funding.}

We have made some use of rotation-age
relationships \citep{2008ApJ...687.1264M}; additional work to determine precise rotation periods would
be useful for specific stars.  Lastly, we have proposed earlier \citep{2015MNRAS.448L..90J} that the solar
$b$ and $y$ colors be monitored by placing an inert sphere in geosynchronous orbit and observing it in the
same way as the stars for the lifetime of the sphere.

\vskip 1cm
\noindent{\bf Acknowledgments}
We are grateful to Louis Boyd for his many years of devotion at Fairborn Observatory.  Without his work,
results such as those presented here would remain out of reach to all.  Giuliana de Toma provided
helpful comments on the manuscript.
G.W.H. acknowledges long-term
support from NASA, NSF, Tennessee State University, and the State of Tennessee through its Centers of
Excellence program. The National Center for Atmospheric Research is funded by the National Science
Foundation.


\bibliographystyle{apj}
\bibliography{ms}

\end{document}